\documentclass{article}

\usepackage[super]{natbib}
\setcitestyle{comma}
\usepackage{PRIMEarxiv}
\usepackage{amsmath}
\usepackage{amssymb}
\usepackage{xcolor}
\usepackage[utf8]{inputenc} % allow utf-8 input
\usepackage[T1]{fontenc}    % use 8-bit T1 fonts
\usepackage{hyperref}       % hyperlinks
\usepackage{url}            % simple URL typesetting
\usepackage{booktabs}       % professional-quality tables
\usepackage{amsfonts}       % blackboard math symbols
\usepackage{nicefrac}       % compact symbols for 1/2, etc.
\usepackage{microtype}      % microtypography
\usepackage{lipsum}
\usepackage{fancyhdr}       % header
\usepackage{graphicx}       % graphics
\graphicspath{{media/}}     % organize your images and other figures under media/ folder

\title{
Transformability reveals the interplay of dynamics across different network orders
}

\author{
  Ming Xie \\
  Department of Control Science and Engineering,\\
  Zhejiang University,\\
   Hangzhou 310027, China\\
  \texttt{mingxie@zju.edu.cn} \\
      \And
  Shibo He \\
  Department of Control Science and Engineering,\\
  Zhejiang University,\\
   Hangzhou 310027, China\\
  \texttt{s18he@zju.edu.cn} \\
   \And
  Aming Li \\
  Center for Systems and Control, College of Engineering,\\
Peking University,\\ 
Beijing 100871, China\\
  \texttt{liaming@pku.edu.cn} \\
  \And
  Zike Zhang \\
  College of Media and International Culture,\\
Zhejiang University,\\ 
Hangzhou 310058, China\\
  \texttt{zkz@zju.edu.cn} \\
     \And
  Youxian Sun \\
  Department of Control Science and Engineering,\\
  Zhejiang University,\\
   Hangzhou 310027, China\\
  \texttt{sunyx@zju.edu.cn} \\
   \And
  Jiming Chen \\
  Department of Control Science and Engineering,\\
  Zhejiang University,\\
   Hangzhou 310027, China\\
  \texttt{cjm@zju.edu.cn} \\
  \And
  \\
  (Dated: January 27, 2025)
}

\begin{document}
\maketitle
\begin{abstract}

Recent studies have investigated various dynamic processes characterizing collective behaviors in real-world systems. However, these dynamics have been studied individually in specific contexts. In this article, we present a holistic analysis framework that bridges the interplays between dynamics across networks of different orders, demonstrating that these processes are not independent but can undergo systematic transformations. 
Focusing on contagion dynamics, we identify and quantify dynamical and structural factors that explains the interplay between dynamics on higher-order and pairwise networks, uncovering a universal model for system instability governed by these factors.
Furthermore, we validate the findings from contagion dynamics to opinion dynamics, highlighting its broader applicability across diverse dynamical processes. Our findings reveal the intrinsic coupling between diverse dynamical processes, providing fresh insights into the distinct role of complex dynamics governed by higher-order interactions.
\end{abstract}

% keywords can be removed
\keywords{Network Dynamics \and Higher-order networks \and Network representations}

\section{Introduction}
\begin{figure}[ht]
  \centering
  \includegraphics[width=\linewidth]{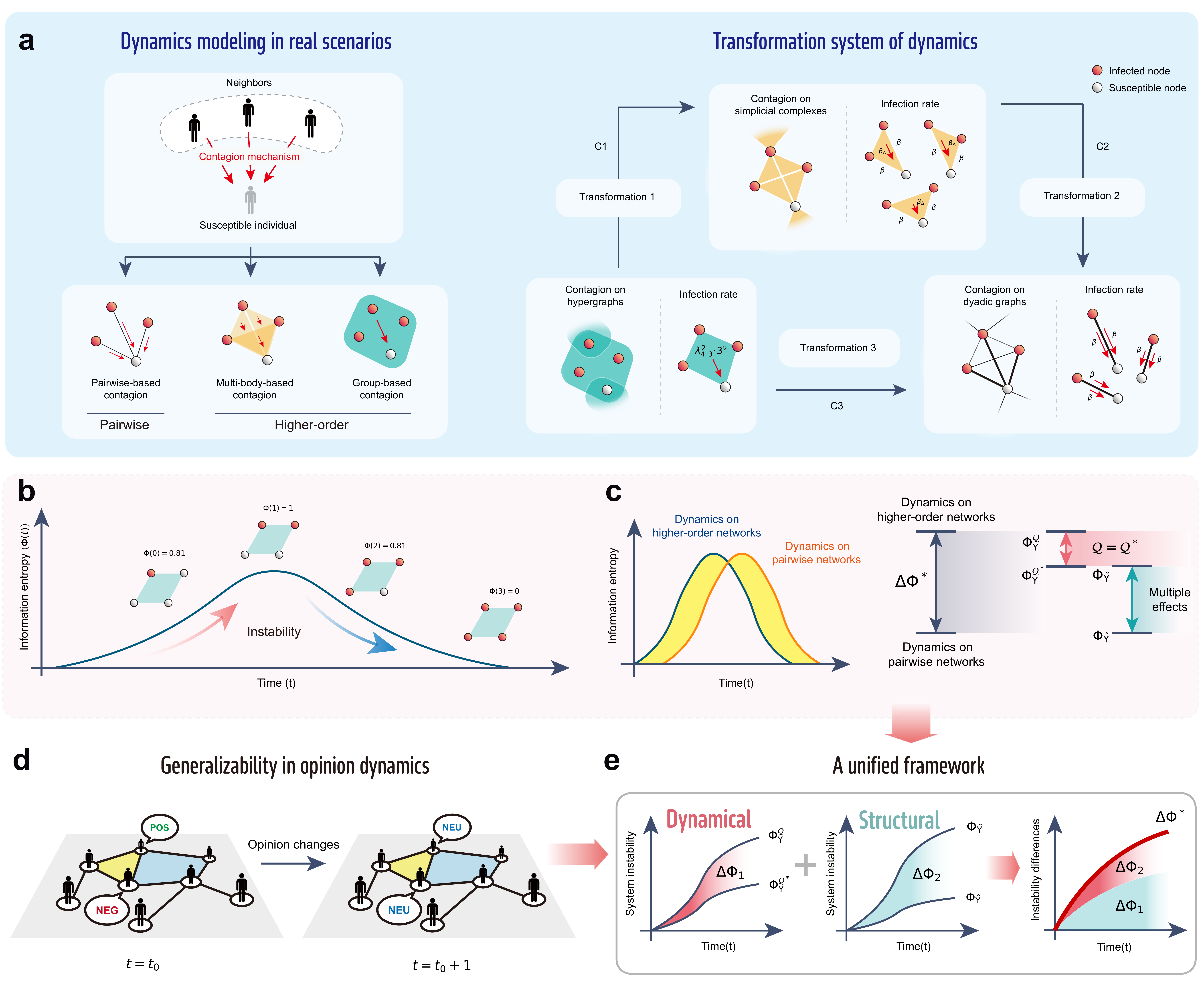}
\caption{\textbf{Overview of the workflow architecture.} 
\textbf{a} Schematic representation (left) of contagion dynamics in disease transmission. The plot (right) depicts the transformation system among SI contagion dynamics modeled on hypergraphs, simplicial complexes, and dyadic graphs, with a maximum hyperedge size of \(M = 4\) and \(\kappa = 2\). Infection rates are represented as \(\beta\) for 1-simplices and \(\beta_\Delta\) for 2-simplices.  
\textbf{b} Temporal evolution of information entropy (\(\Phi\)) during contagion processes.  
\textbf{c} The interplay between dynamics on the higher-order network (\(\Upsilon\)) and its pairwise counterpart (\(\hat{\Upsilon}\)). The left panel compares the dynamics on the two network structures, quantified via changes in information entropy. The right panel disentangles the contributions of nonlinearity (\(\mathcal{Q}\)) and higher-order structures to the observed dynamical differences.  
\textbf{d} Applicability of the framework to opinion dynamics. The plot illustrates how individuals from distinct discussion groups (yellow and blue) adjust their positive (POS) and negative (NEG) opinions to a neutral (NEU) state as a result of mutual influence during interaction.  
\textbf{e} A unified framework demonstrating the interplay of dynamics across different orders (\(\Delta \Phi^*\)) as governed by the combined contributions of dynamical (\(\Delta \Phi_1\)) and structural (\(\Delta \Phi_2\)) factors.
}
  \label{fig1:framework}
\end{figure}

Characterizing contagion dynamics is pivotal to encoding the evolution of collective behaviors~\cite{ murphy2021deep, li2021dynamics, huang2024identifying}, and has wide applications spanning pathogen transmission~\cite{sun2021competition}, information diffusion~\cite{yu2021modeling}, cultural socialization~\cite{yan2024emotion}, etc~\cite{nazerian2024efficiency, de2016physics}. Existing contagion dynamic models, such as the susceptible-infected (SI) model~\cite{allen1994some} and threshold models, have illuminated essential dynamical features, including phase transitions~\cite{qu2017sis, xie2021detecting} and endemic outbreaks~\cite{juher2013outbreak, de2024determinants}. These paradigms, however, are largely based on pairwise interactions (Fig.~\ref{fig1:framework}a), where the state of a node is dictated by interactions between individual neighbors.

The prevalence of group interactions in real-world systems has led to the emergence of higher-order networks~\cite{boccaletti2023structure, horsevad2022transition, malizia2025hyperedge}, examplied by hypergraphs~\cite{contisciani2022inference, badalyan2024structure} and simplicial complexes~\cite{gambuzza2021stability, ghorbanchian2021higher}, as foundational tools for studying contagion dynamics~\cite{lee2021hyperedges, battiston2020networks, zhang2023higher}, which overcome the limitations of traditional models in the following aspects.
First, higher-order networks effectively represent multifaceted interactions among nodes (i.e., higher-order interactions)~\cite{barash2012critical, barrat2022social}, allowing for diverse propagation pathways~\cite{iacopini2019simplicial, centola2007complex}. 
Second, these network structures reveal unique dynamical properties, e.g., discontinuous phase transitions and bistable states,  offering fresh perspectives in characterizing complex contagion processes~\cite{st2022influential,cencetti2023distinguishing}.

While substantial efforts have been made to understand the dynamics of contagion, ranging from pairwise-based mechanisms to those driven by higher-order interactions~\cite{chowdhary2021simplicial, guzzo2017malevolent, de2020social} (see Fig.~\ref{fig1:framework}a), these studies have largely overlooked the interplay between dynamical processes over networks of different orders. 
Notably, finding such an interplay is fundamental to explain evolutionary principles that govern chaotic dynamics in empirical systems~\cite{st2024nonlinear} and uncover driving factors that enable understanding dynamical processes through a more nuanced and complex lens, thereby providing specific strategies to control contagion. 
Unfortunately, it is  
challenging to doing so due to the following two reasons.  
First,  dynamics models are application defined,  and often depend on domain-specific rules, creating pronounced structural and mechanistic disparities. Such heterogeneity poses formidable obstacles to the establishment of a unified theoretical framework capable of encompassing the full spectrum of dynamic phenomena.  
Second, modeling dynamics on higher-order networks necessitates the explicit incorporation of multi-scale interactions and the intricate nonlinear processes underpinning these mechanisms. This intrinsic complexity not only deepens the theoretical challenges but also amplifies computational demands.

In this article, we take the first attempt to address such a challenging issue. We first develop a framework where contagion dynamics are transformable across different network orders, as illustrated in Fig.~\ref{fig1:framework}a. 
We reveal that different dynamics are not mutually independent but can be transformed under certain conditions. 
Secondly, as the dynamical process inherently reflects the variations of node states within a system, which can be characterized through the evolution of system instability, we naturally leverage information entropy to globally quantify the evolution of instability generated throughout these contagion processes (Fig.~\ref{fig1:framework}b).
This approach further facilitates the identification of factors driving the disparities in dynamics across network orders, particularly from pairwise to higher-order networks, see Fig.~\ref{fig1:framework}c.
We show that the structural (i.e., the presence of multiple contagion contacts between nodes embedded within higher-order interactions) and dynamical (i.e., the nonlinear impact induced by contagion processes mediated by higher-order interactions) aspects serve as two independent factors contributing to the underlying transformation between the dynamics on the networks of different orders. This finding allows us to propose a unified framework to capture this transformation.
Finally, we validate the generalizability of our findings within the realm of opinion dynamics~\cite{tang2021exchange, ferraz2023multistability, schawe2022higher}, showcasing its robustness and applicability across a broad spectrum of dynamical systems (Fig.~\ref{fig1:framework}d). 
To the best of our knowledge, this study is the first to provide a holistic transformation analysis for dynamics across network orders. In addition, we offer a fundamental understanding of the roles of both structural and dynamical factors in dynamics across network orders from a perspective of system instability,
thereby providing insights into the distinctive role of higher-order dynamical systems.

\section{Results}
\subsection{Transformability across diverse contagion dynamics}

To explore the transformability among contagion dynamics across network orders, we consider the SI contagion on hypergraphs (\(\mathcal{H}\)) , simplicial complexes (\(\mathcal{C} \)) and dyadic graphs ($\mathcal{G}$), whose representations are aligned through network projection, as described in Methods.

For the contagion dynamics on hypergraphs, we isolate the effects of higher-order interactions by defining \(\lambda_{m,\chi} \cdot \chi^\nu\) as the infection rate within an \(m\)-sized hyperedge containing \(\chi\) nodes in the infected (I) state. Accordingly, the infection probability of a susceptible node \(i\) under  SI model in \(\mathcal{H}\), denoted as \(P_{\mathcal{H}}^i(t)\) (see detailed description in the Methods section), can be refined as follows:
\begin{eqnarray}
\label{orginal_infection_rate}
    P_{\mathcal{H}}^i(t)=1-\prod_{m=2}^M\prod_{\chi=1}^{m-1}(1-\lambda^{\kappa}_{m, \chi} \cdot \chi^\nu \cdot A_h^m).
\end{eqnarray}
Here,  \( M \) denotes the maximum size of hyperedges, while \(\kappa\) indicates the mapping ratio.
Similarly, infection probabilities of node $i$ under the dynamics on the simplicial complex and dyadic graph are denoted as $P_{\mathcal{C}}^i(t)$ and $P_{\mathcal{G}}^i(t)$, respectively. (see Methods for details)

We then determine the transformable conditions and show that the SI process on any hypergraph, with hyperedges of arbitrary size $m$ and an arbitrary mapping ratio \(\kappa\), can be consistent with contagion process on the simplicial complex representation, as provided by the following condition (Condition 1):
\begin{eqnarray}
\label{eqation6}
        \lambda^{\kappa}_{m, \chi} = \frac{1
        }{\chi^\nu} 
        \sum_{i=\tau}^{\tau^*} \binom{\chi}{i} \binom{m-\chi-1}{\kappa-i} \sum_{j=1}^{i} \binom{i}{j} \beta_j,
\end{eqnarray}
where $\tau=\max(1, \kappa+1-(m-\chi))$ and $\tau^*=\min(\kappa, \chi)$. Further details regarding the transformation can be found in the Supplementary Information.
Combining Eq.~\ref{eqation6} with the condition which defined as \(\beta_{\delta} = 0(\delta > 1)\) (Condition 2), the contagion dynamics on \(\mathcal{H}\) become equivalent to those on its projected dyadic graph with multiple edges when the following condition (Condition 3) is satisfied:
\begin{equation}
\label{eqation5}
    % \lambda_1=\lambda_{m-1}=\lambda_{\chi_{1, \cdots, m-2}}=\beta \neq 0.
     % \lambda_{\chi_{1, \cdots, \chi-2}}=(\chi-3)\cdot\beta,
      % \lambda^\kappa_{m, \chi}=\frac{1 }{\chi^{\nu}} \binom{\chi}{\tau}\cdot\tau\beta.
      \lambda^\kappa_{m, \chi}= \frac{1
    }{\chi^\nu} 
     \sum_{i=\tau}^{\tau^*} i \cdot \binom{\chi}{i} \binom{m-\chi-1}{\kappa-i} \beta .
\end{equation}
Take a hypergraph $\mathcal{H}_3$ as an example, where the maximum size of hyperedge is 3 (i.e., $M=3$), and the mapping ratio is given by \(\kappa = 2\). 
The contagion dynamics on a hypergraph align with those on its corresponding simplicial complex when the infection probabilities \(P_{\mathcal{H}_3}^i(t)\) and \(P_{\mathcal{C}}^i(t)\) coincide. This equivalence is achieved under the conditions \(\lambda^2_{2,1} \cdot 1^\nu = \lambda^2_{3,1} \cdot 1^\nu = \beta \neq 0\) and \(\lambda^2_{3,2} = \frac{2\beta + \beta_\Delta}{2^\nu}\) (C1).
Furthermore, the complex contagion dynamics on the simplicial complex can be reduced to a simple contagion on the projected multi-edge graphs when \(\beta_{\Delta} = 0\) (C2).
Additionally, the probability \(P^i_{\mathcal{H}_3}(t)\) matches \(P^i_{\mathcal{G}}(t)\) when both C1 and C2 are satisfied (C3).
\begin{figure}[tp]
  \centering
  % \fbox{\rule[-.5cm]{4cm}{4cm} \rule[-.5cm]{4cm}{0cm}}
  \includegraphics[width=\linewidth]{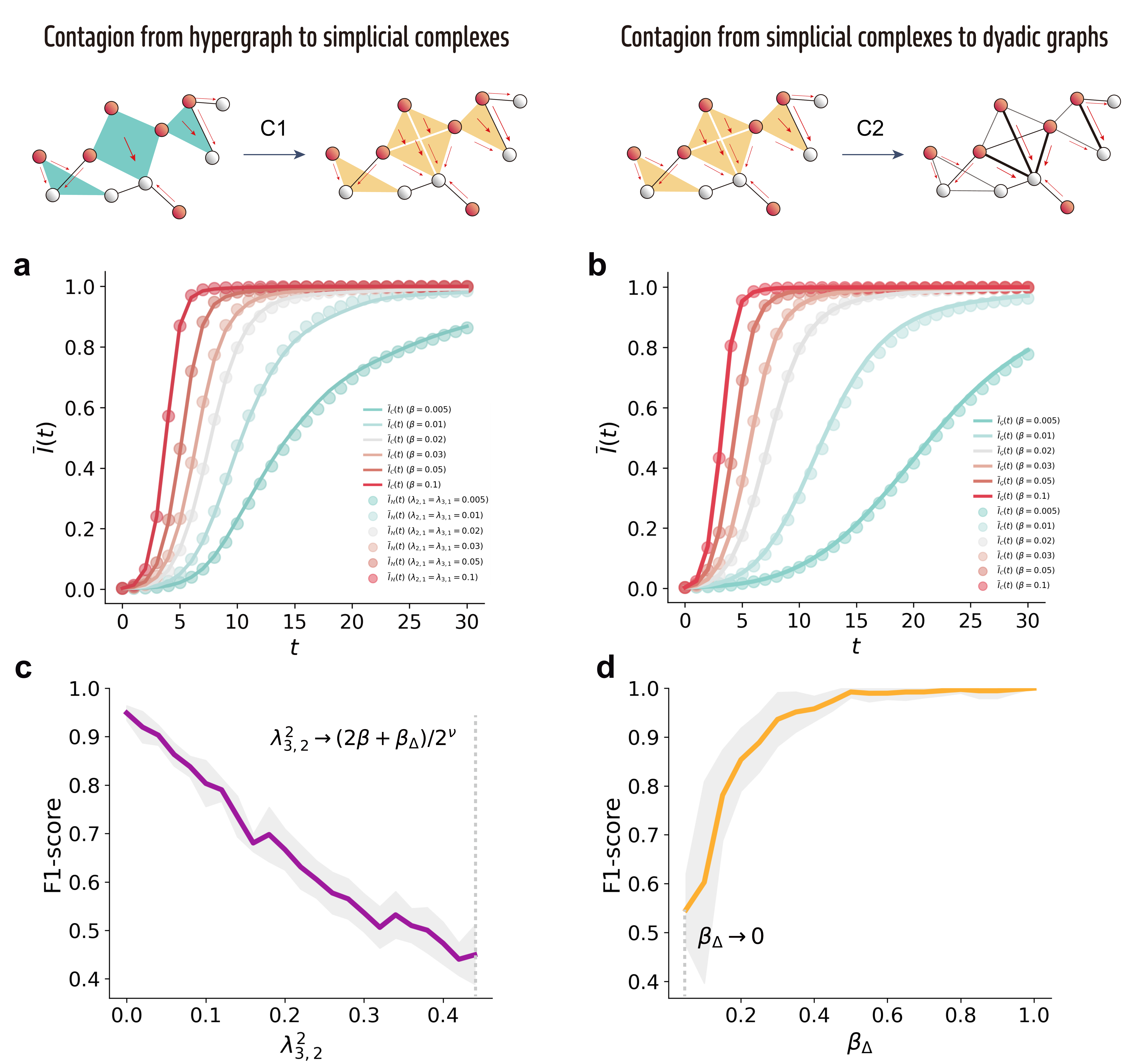}
\caption{
\textbf{The interplay between dynamic processes.}
\textbf{a} Average infection density \(\overline{I}(t)\) for contagion dynamics on a hypergraph and its projected simplicial complex, based on an empirical high-school social network. For the hypergraph (dots), \(\lambda_{2,1} = \lambda_{3,1} = \beta\), \(\lambda_{3,2} = (2\beta + \beta_\Delta) / 2^\nu\); for the simplicial complex (lines), \(\beta_\Delta = 0.5\). \(\beta\) varies from \(0.005\) to \(0.1\), with \(\kappa = 2\).  
\textbf{b} Comparison of \(\overline{I}(t)\) for dynamics on simplicial complex (dots) and those on projected dyadic graph (lines). Results are averaged over 500 simulations with \(\mathcal{T} = 30\).  
\textbf{c, d} F1-score variation as the system approaches critical states C1 (\textbf{c}) and C2 (\textbf{d}). Lower F1-scores in \textbf{c} indicate reduced accuracy in distinguishing hypergraph-based contagion, while higher F1-scores in \textbf{d} reflect improved classification of simplicial complex dynamics. The SVM classifier uses node degree and infection order features from simulations with \(\lambda = 0.04, \nu = 1\) (hypergraphs) and \(\beta = 0.04, \beta_\Delta = 0.8\) (simplicial complexes). Results are averaged over \(10^3\) runs with \(\mathcal{T} = 50\). 
}
\label{fig2:Interrelationships}
\end{figure}

To verify our proposed transformation system across dynamics, we conduct simulations of contagion dynamics on both hypergraphs and simplicial complexes, along with their projections. 
We monitor the evolution of the infection density \(\overline{I}(t)\) over time for each simulation, as depicted in Fig.~\ref{fig2:Interrelationships}a and ~\ref{fig2:Interrelationships}b. 
We observe that, under the first set of condition (C1), \(\overline{I}_{\mathcal{H}}(t)\) approximately converges to the curves of \(\overline{I}_{\mathcal{C}}(t)\). Building on this, when the additional condition C2 is imposed (satisfying C3), the curve further aligns with \(\overline{I}_{\mathcal{G}}(t)\).
These findings suggest that the contagion dynamics on hypergraphs can effectively be transformed to those on simplicial complexes and, subsequently, to those on projected dyadic graphs with multiple edges. 

In addition, to numerically test the progressive effect of proposed conditions on the transformation between contagion dynamics, we classify dynamics using a model of Support Vector Machine (SVM). The classification is based on the correlation between the node degree and infection order~\cite{cencetti2023distinguishing}, with the results presented in Fig.~\ref{fig2:Interrelationships}c and ~\ref{fig2:Interrelationships}d. We observe that as C1 is progressively satisfied, the classification performance decreases with a lower F1-score, making it more difficult to distinguish between dynamics on hypergraphs and simplicial complexes.
In contrast, when C2 is less restrictive, classification improves between dynamics on simplicial complexes and dyadic graphs, as evidenced by the increased F1-score. 
These results suggest that when the transformable condition between two contagion dynamics becomes more restrictive, the dynamics tend to converge; otherwise, they diverge further, underscoring the complex interplay between the two processes.

To sum up, the proposed transformation system bridges the intricate relationships across diverse dynamics by showing that diverse contagion dynamics are interconnected rather than isolated.
It also suggests that dynamics on hypergraphs inherently exhibit the characteristics of those on both dyadic graphs and simplicial complexes.

\subsection{A unified framework from the perspective of system instability}
Building on the transformation  among dynamics via the infection probability, we further explore factors driving the transitions across network orders, i.e., transformations between higher-order and pairwise networks. 
To this end, we advance from a localized perspective focusing on node infection probabilities to a holistic perspective, capturing the systemic evolution of node states at the global scale.
We employ information entropy to characterize the evolution of dynamics, as it effectively captures system instability arising from the diversity of node states within the network during the dynamic process, as shown in Fig.~\ref{fig1:framework}b. 
Leveraging entropy, we identify and quantify the key factors governing dynamic transformations, constructing a comprehensive framework to unravel the dynamic distinctions between higher-order and pairwise networks.

\textit{Identifying contributing factors of dynamic transformations across network orders.} 
To explore the contributing factors, we first obtain the curves of information entropy \(\Phi\) from the infection density \(\overline{I}(t)\) for contagion dynamics on both higher-order and pairwise networks, as illustrated in the first plot of Fig. 1c (see detailed information in the Supplementary Information), and then calculate the information differences between these two curves (represented as $\Delta \Phi^*$ in grey in the second plot of Fig. 1c).
We identify that the contributing factors can be decomposed into the following two distinct components: 
(i) a dynamical factor, stemming from the nonlinearity during contagion processes; and (ii) a structural factor, attributed to higher-order interactions that manifest in multiple contacts. The effects of dynamical and structural factors are characterized as dynamical uncertainty ($\Delta \Phi_1$ in Fig. 1c) and structural uncertainty ($\Delta \Phi_2$ in Fig. 1c), respectively.

To explicitly characterize the nonlinearity inherent in node infections, we define a measure, \(\mathcal{Q}\), as follows: 
\begin{align}
    &\mathcal{Q} = \mathcal{A} \cdot \mathbb{I}_{\Upsilon=\mathcal{H}} + \mathcal{B} \cdot \mathbb{I}_{\Upsilon=\mathcal{C}},
\end{align}
where \( \mathcal{A}= \left(\lambda^\kappa_{m, \chi}- \frac{1
    }{\chi^\nu} 
     \sum_{i=\tau}^{\tau^*} i \cdot \binom{\chi}{i} \binom{m-\chi-1}{\kappa-i} \beta \right)^2\) and \( \mathcal{B} = \beta_{\delta}^2 (\delta > 1) \). Here, \(\mathbb{I}\) is an indicator function, where \(\mathbb{I}_{\Upsilon=\mathcal{H}}\) equals \(1\) if the system \(\Upsilon\) is a hypergraph and \(0\) if it is a simplicial complex.
This measure indicates the degree of restrictiveness associated with the proposed transformable conditions. Specifically, when C2 (or C3) is satisfied for dynamics on a hypergraph (or a simplicial complex), the measure reduces to \(\mathcal{Q}= 0\). Notably, in such cases, we denote \(\mathcal{Q}\) as \(\mathcal{Q}^*\) for clarity.

For a structural factor, we observe that it is captured by the differences in interaction structures, which are intrinsic to networks of different orders.  Specifically, these differences arise from the presence of multiple contacts between nodes in higher-order networks compared to the single contacts observed between nodes in pairwise networks, which are reflected in the edge disparities between two projections of higher-order networks—namely, the multi-edge graph \(\tilde{\mathcal{H}}\) and the simple graph \(\hat{\mathcal{H}}\).

\begin{figure}
  \centering
  \includegraphics[scale=0.6]{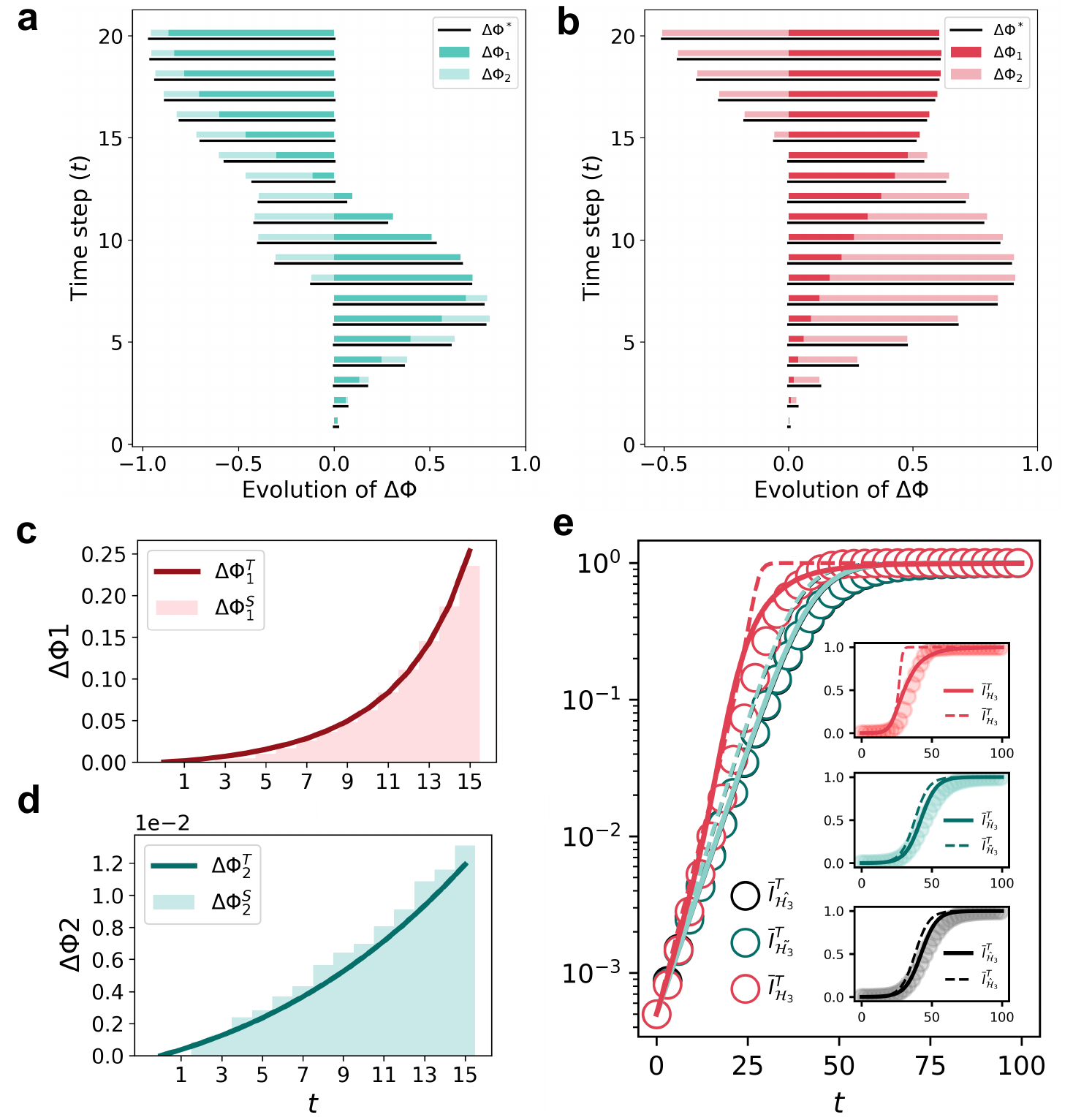}
  \caption{\textbf{Identification and quantification of contributing factors in contagion dynamics.} 
  \textbf{a}-\textbf{b} Temporal evaluation of \(\Delta \Phi_1(t)\) and \(\Delta \Phi_2(t)\) in SI contagion processes simulated on real-world datasets of primary school and high school interactions. In \textbf{a}, \(\Delta \Phi_1(t)\) is calculated as \(\Delta \Phi_1(t) = \Phi^{\mathcal{Q}}(t) - \Phi^{\mathcal{Q}^*}(t)\), with \(\mathcal{Q}(\lambda = 0.005, \nu = 4)\) and \(\mathcal{Q}^*(\nu = 1)\), where \(\lambda\) satisfies Eq.~\ref{eqation5} for \(\mathcal{Q}^*\). \(\Delta \Phi_2(t)\) is derived as \(\Phi_{\tilde{\mathcal{H}}}(t) - \Phi_{\hat{\mathcal{H}}}(t)\).  
  In \textbf{b}, for the simplicial complex \(\mathcal{C}\), \(\Delta \Phi_1(t)\) is evaluated by setting \(\mathcal{Q}(\beta_\Delta = 0.9)\) and \(\mathcal{Q}^*(\beta_\Delta = 0)\), while \(\Delta \Phi_2(t)\) is calculated similarly to \textbf{a}. The total entropy difference, \(\Delta \Phi^*\) (dotted line), closely matches the sum of \(\Delta \Phi_1(t)\) (dark-colored bars) and \(\Delta \Phi_2(t)\) (light-colored bars) over time. Results are averaged over 500 independent simulations.  
  \textbf{c}-\textbf{d} Comparison of theoretical predictions (\(\Delta \Phi^T\), solid lines) from the SI model with simulation results (\(\Delta \Phi^S\), bars) for \(\Delta \Phi_1(t)\) and \(\Delta \Phi_2(t)\). Strong agreement is observed between theory and simulation.  
  \textbf{e} Comparison of quantitative evaluation of the mean infection density \(\overline{I}(t)\) between the original and the refined models. Theoretical predictions \(\overline{I}^T(t)\) (solid lines) align closely with simulation results \(\overline{I}^S(t)\) (dots), as shown for simplicial complexes \(\overline{I}_{\mathcal{H}_3}(t)\), projected multi-edge graphs \(\overline{I}_{\tilde{\mathcal{H}_3}}(t)\), and simple graphs \(\overline{I}_{\hat{\mathcal{H}_3}}(t)\). The inset presents the non-logarithmic results of \(\overline{I}(t)\) with error bars. Simulations utilize the LSODA algorithm to numerically solve \(\overline{I}_{\mathcal{H}_3}(t)\). Results are averaged over 500 numerical realizations per data point.
}
  \label{fig5:experiments}
\end{figure} 

We conduct the validation on both hypergraphs and simplicial complexes generated from empirical social contact systems, aiming to confirm that the differences of entropy in dynamics between higher-order networks and their projected simple graphs can indeed be decomposed into dynamical and structural components.
Fig.~\ref{fig5:experiments}a shows the results of the dynamics examination on a hypergraph.  We first evaluate the difference of information entropy \(\Delta \Phi^*\) by conducting the simulation of contagion dynamics on both the hypergraph $\mathcal{H}$ and its projected simple graph $\hat{\mathcal{H}}$. 
Subsequently, we simulate the contagion process in $\mathcal{H}$ for both \(\mathcal{Q}\neq\mathcal{Q}^*\) and \(\mathcal{Q}=\mathcal{Q}^*\) to determine the entropy difference \(\Delta \Phi_1\) resulting from dynamical factor, where  \(\Delta \Phi_1\) is obtained by $\Phi_{\mathcal{H}}^{\mathcal{Q}} - \Phi_{\mathcal{H}}^{\mathcal{Q}^*}$.
Furthermore, we obtain the entropy difference \(\Delta \Phi_2\) between the contagion dynamics on the projection \(\tilde{\mathcal{H}}\) and \(\hat{\mathcal{H}}\) (i.e., \(\Delta \Phi_2 = \Phi_{\tilde{\mathcal{H}}} - \Phi_{\hat{\mathcal{H}}}\)), aiming to assess the structural uncertainty.
As shown in Fig.~\ref{fig5:experiments}a, the total information difference \(\Delta \Phi^*\) shows a trend that aligns approximately with the sum of \(\Delta \Phi_1\) and \(\Delta \Phi_2\). Fig.~\ref{fig5:experiments}b presents the validation performed on a simplicial complex \(\mathcal{C}\), where we observe consistent results. 
Moreover, we provide detailed theoretical proofs for this finding, that is, \(\Delta \Phi^*\) is equivalent to \( \Phi^{\mathcal{Q}}_\mathcal{C}(t) - \Phi^{\mathcal{Q}^*}_\mathcal{C}(t) + \Phi_{\tilde{\mathcal{C}}}(t) - \Phi_{\hat{\mathcal{C}}}(t)\), with the specific steps elaborated in the Supplementary Information. 
This suggests that, while structural and dynamical contributions  are additive, driving a potential dynamic transformation across network orders from higher-order to pairwise networks.

\textit{Quantifying contributing factors.} Given that both structural and dynamical factors jointly contribute to the variations in dynamics across different network orders, we rigorously quantify their respective contributions using the mean-field theory for $\overline{I}(t)$.
Specifically, we derive the master equations governing the contagion dynamics on a simplicial complex and its projected networks (see Supplementary Information for details). 
Fig.~\ref{fig5:experiments}c-d illustrates the comparison between the theoretical solutions and simulation results for \(\Delta \Phi_1\) and \(\Delta \Phi_2\).
We generate networks with specific topologies, including node degree~\cite{de2020social} and 2-simplex degree~\cite{iacopini2019simplicial}, through a network configuration model~\cite{iacopini2019simplicial}, with procedures outlined in the Supplementary Infomation. 
The results demonstrate that, under the specified parameters, our quantitative model effectively captures the information differences caused by each of the two impacts.
Besides, we test the performance of the quantitative model across various infection rates (\(\beta\) and \(\beta_\Delta\)), recovery rates \(\mu\), network densities \(d\), and network sizes \(N\). 
Detailed findings are included in the Supplementary Information, indicating that network size significantly influences the accuracy of information discrepancy quantification in the model.
In the Supplementary Information, the additional results for the SIS and SIR models are also presented, which yields similar outcomes. 

However, we observe that our theoretical model can occasionally yield less accurate quantification across certain parameter ranges, due to its reliance on uniform average node degrees and average 2-simplex degrees. This limitation arises from the presence of highly central nodes alongside low-degree peripheral (leaf) nodes in networks. 
To better address the heterogeneity in node topology, we introduce the concept of effective degree and develop a tailored differential equation (see Supplementary Information for a detailed description).
The refined theoretical model integrates the distinctive topological features of the network, accounting for the distributions of node degree, 2-simplex degree, and effective degree.
Fig.~\ref{fig5:experiments}e presents the results from the refined quantification, which incorporates differential equations and fine-grained nodes' topologies alongside simulation results, demonstrating significant improvement in quantification accuracy.

\textit{A unified framework.}
Since we have identified the factors contributing to the differences between higher-order and pairwise networks, and have provided an efficient quantification for them, we extend this finding to accommodate a broader range of dynamic categories.
This conclusion is expressed in the following form:
\begin{equation}
\label{univseral_eq}
    \Delta \Phi^* = \Delta \Phi_1 + \Delta \Phi_2 = 
    \Phi^{\mathcal{Q}}_{\Upsilon}(\mathcal{F}) - \Phi^{\mathcal{Q}^*}_{\Upsilon}(\mathcal{F}) + \Phi_{\tilde{\Upsilon}}(\mathcal{F}) - \Phi_{\hat{\Upsilon}}(\mathcal{F}).
\end{equation}
In this context, \(\Phi\) indicates a measure captures the system instability. \(\Upsilon\), \(\tilde{\Upsilon}\), and \(\hat{\Upsilon}\) represent a higher-order network, its multi-edge projection, and its simple projection, respectively. 
The term \(\mathcal{Q}\) denotes nonlinear factors from higher-order interactions, while \(\mathcal{Q}^*\) represents their transformed linear equivalents.
The function \(\mathcal{F}=f(\sigma_i(t))\) refers to the state integration operation, with \(\sigma_i(t)\) (\(i \in V\)) denotes the state of each node \(i\) within the network. Here, \(\sigma_i(t) = 1\) indicates that node \(i\) is in the activated state, while \(\sigma_i(t) = 0\) otherwise. Specifically, during the contagion dynamics, we express \( f(\sigma_i(t))=\overline{I}(t)\), where \(f(\sigma_i(t)) = \sum_{i \in V_{\textbf{I}}} \sigma_i(t)\), with $V_{\textbf{I}}$ denoting the set of infected nodes within the network. 
At this point, we present a unified framework that interprets the underlying mechanisms driving the transformation between dynamics on higher-order and pairwise networks.
Besides, we note that the applicability of this framework is contingent upon the condition that  \(\mathcal{Q}^*\) exists. 
Through this framework, we disentangle the interplay of dynamics across different network orders into two explicit and additive components, namely dynamical and structural, thereby enabling the framework's adaptability to a broad range of processes, including contagion and opinion dynamics.

\begin{figure}[ht]
  \centering
  \includegraphics[width=\linewidth]{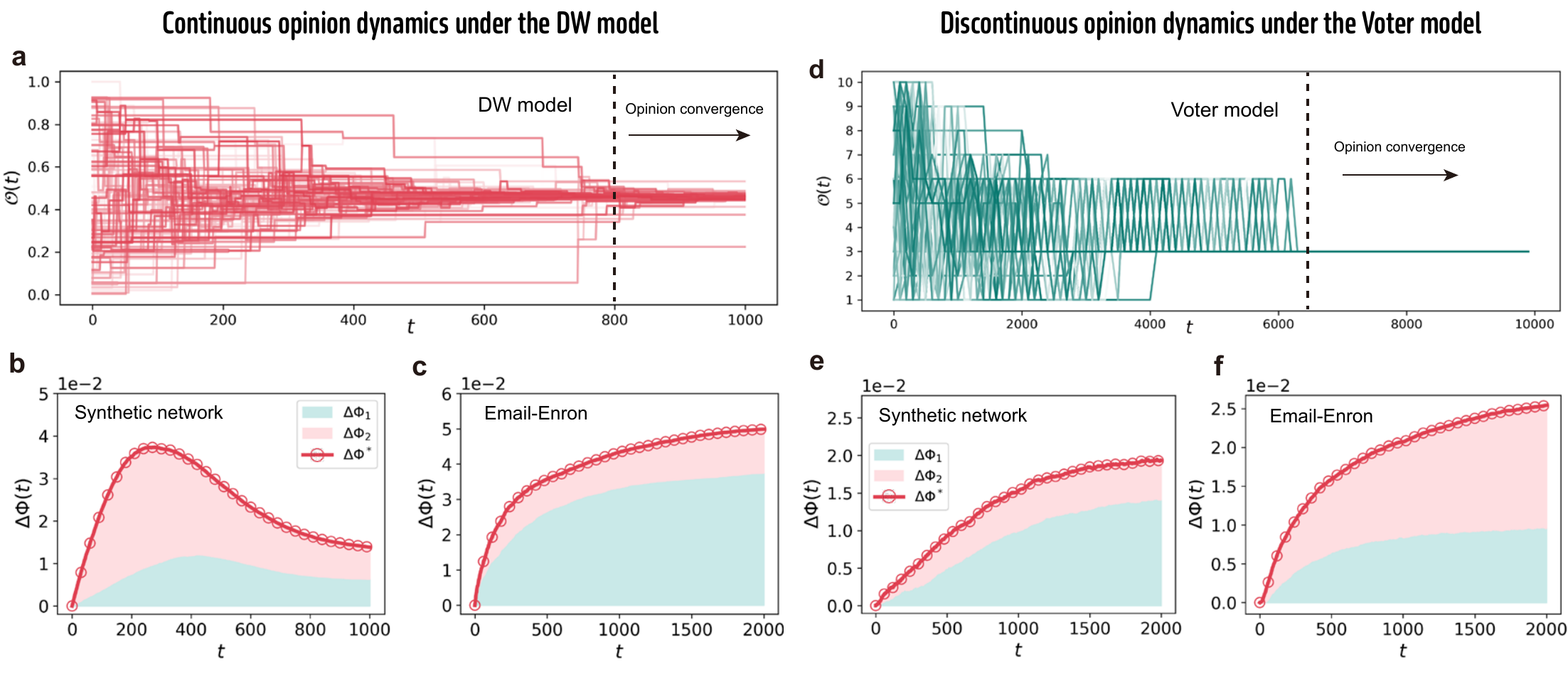}
  \caption{
  \textbf{Validations in DW and Voter opinion dynamics.}
  \(\textbf{a}\)  Opinion variations of nodes in $\mathcal{H}$ in a simulation of DW model. The initial opinion values are set to follow a uniform distribution. As time progresses, continuous opinions among nodes evolve and eventually converge to a steady state. \(\textbf{b-c}\) 
  The relationships among the dynamical uncertainty \( \Delta \Phi_1\), the structural uncertainty \( \Delta \Phi_2 \), and the total differences of system instability (\( \Delta \Phi^* \)) underlie the interplay between higher-order and pairwise networks in both synthetic and empirical systems. 
  \(\textbf{d}\)  Opinion variations of nodes in $\mathcal{H}$ in a simulation of Voter model. The initial opinion values are distributed uniformly. Over time, discontinuous opinions among nodes evolve and eventually converge to a steady state.
  \(\textbf{e-f}\) 
  The relationships among the dynamical uncertainty \( \Delta \Phi_1\), the structural uncertainty \( \Delta \Phi_2 \), and the total differences of system instability (\( \Delta \Phi^* \)) underlie the interplay between higher-order and pairwise networks in both synthetic and empirical systems. Each value of \( \Phi(t) \) is based on $10^3$ simulations.
  }
  \label{fig7:experiments}
\end{figure}

\subsection{Generalizability in opinion dynamics}
We have focused  on contagion dynamics thus far. Next, we evaluate the generalizability of the proposed framework beyond the scope of contagion processes.
Opinion dynamics models study how opinions evolve, spread, and stabilize in  networks. 
Take a social network of friends debating their coffee preference for an example, illustrated in Fig.~\ref{fig1:framework}e. Individuals begin with one of three opinions: positive (POS), negative (NEG), or neutral (NEU), reflecting enthusiasm, aversion, or indifference. These opinions evolve as individuals influence one another through their connections. In this case, a coffee enthusiast might convert an opponent or moderate their stance, shifting to neutrality. 
Building on the proposed unified framework, we validate it in both the continuous and discontinuous opinion dynamics, namely, DW model~\cite{dong2018survey, das2014modeling} and Voter model~\cite{mori2019voter, shi2013multiopinion}, introduced in Methods section. 

\begin{figure}[ht]
  \centering
  \includegraphics[width=\linewidth]{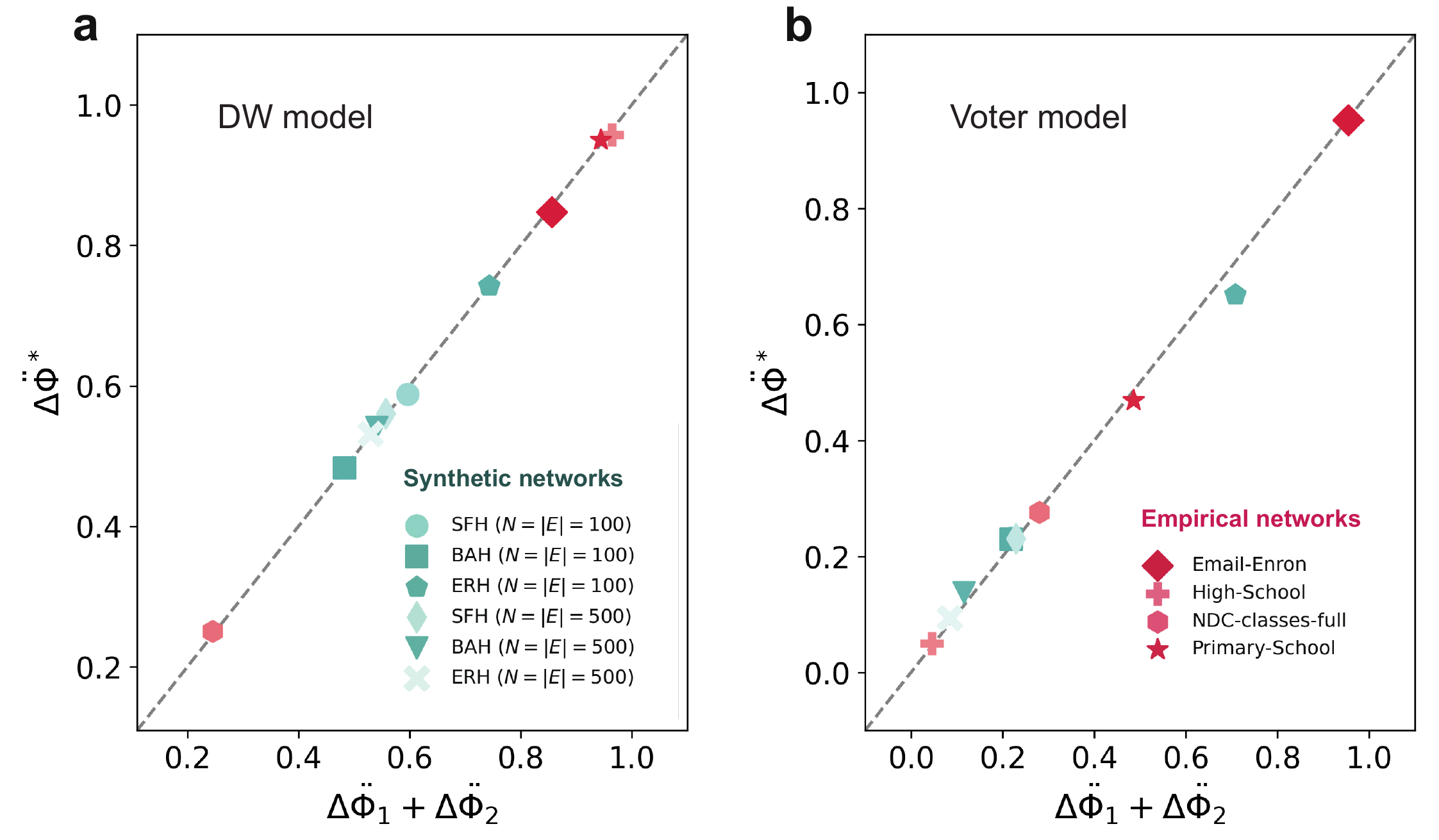}
  \caption{
\textbf{Generalizability of the unified framework.}
\textbf{a} Comparison of \( \Delta \ddot{\Phi}^* \) and \( \Delta \ddot{\Phi}_1 + \Delta \ddot{\Phi}_2 \) in DW dynamics within system \( \Upsilon \) across both synthetic and empirical systems. Synthetic hypergraphs with varying topologies are generated using HyperCL, where the degree distribution exponents are set to \( 2 \), \( 3 \), and \( 5 \) for scale-free hypergraphs (SFH), Barabási–Albert hypergraphs (BAH), and Erdős–Rényi hypergraphs (ERH), respectively, transitioning from scale-free to uniform degree distributions. Empirical systems, including Email-Enron, High-School, NDC-classes-full, and Primary-School networks, are analyzed, with detailed descriptions provided in the Supplementary Information.  
\textbf{b} Comparison of \( \Delta \ddot{\Phi}^* \) and \( \Delta \ddot{\Phi}_1 + \Delta \ddot{\Phi}_2 \) in Voter dynamics within \( \Upsilon \) for both synthetic and empirical systems. Each value is averaged over \( 500 \) simulation runs.
}
  \label{fig8:experiments}
\end{figure}

Considering the differences in dynamical mechanisms between the contagion and opinion dynamics, we need to adopt the proposed framework  to accommodate the characteristics of opinion dynamics. Firstly, the $\mathcal{F}$ involved in the framework of Eq.~\ref{univseral_eq} is denoted as $\mathcal{F}=\mathcal{O}(t)$, where  $\mathcal{O}(t)$ is a set of the opinion values of all nodes, i.e., $\mathcal{O}(t)=\bigcup_i o_i(t)$. Then, the system instability is calculated by the variance of the opinion values across all nodes, i.e., \(\Phi = \mathcal{D}(\mathcal{O}(t))\), where $\mathcal{D}$ is a variance function. In this context, $\mathcal{Q}\neq \mathcal{Q}^*$ for $\alpha < 1$, and $\mathcal{Q}=\mathcal{Q}^*$ when $\alpha=1$ (denoted as $\alpha_1$). Lastly, the framework is performed on the original hypergraph $\mathcal{H}$ into its projections, $\tilde{\mathcal{H}}$ and $\hat{\mathcal{H}}$. 

Next, we validate that the transformation between dynamics on higher-order and pairwise networks reflects the additivity of both dynamical and structural contributions in opinion dynamics.

The validation is initially performed using DW model. 
Figure~\ref{fig7:experiments}a illustrates the evolution of opinion variance for hypergraphs generated by the HyperCL configuration model~\cite{lee2021hyperedges} and their projected networks, showing a steady decline in opinion dispersion under DW opinion dynamics as node opinions gradually converge with increasing \(t\).
Subsequently, we separately calculate the impact arising from the dynamical and the structural contributions using Eq.~\ref{univseral_eq}, defined as \(\Delta \Phi_1 = \mathcal{D}_{\mathcal{H}}^{\alpha}(t) - \mathcal{D}_{\mathcal{H}}^{\alpha_1}(t)\) and \(\Delta \Phi_2 = \mathcal{D}_{\tilde{\mathcal{H}}}(t) - \mathcal{D}_{\hat{\mathcal{H}}}(t)\), where \(\mathcal{D}_{\mathcal{H}}^{\alpha}(t)\) and \(\mathcal{D}_{\mathcal{H}}^{\alpha_1}(t)\) denote the opinion dispersion at time \(t\) from DW model simulations with \(\alpha < 1\) and \(\alpha = 1\), respectively, while \(\mathcal{D}_{\tilde{\mathcal{H}}}\) and \(\mathcal{D}_{\hat{\mathcal{H}}}\) represent the evolving opinion dispersion on \(\tilde{\mathcal{H}}\) and \(\hat{\mathcal{H}}\), respectively. We then compare the sum of the   \(\Delta \Phi_1 \) and \(\Delta \Phi_2\) with the total difference \(\Delta \Phi^* = \mathcal{D}_{\mathcal{H}}^{\alpha}(t) - \mathcal{D}_{\hat{\mathcal{H}}}(t)\), as shown in Fig.~\ref{fig7:experiments}b and ~\ref{fig7:experiments}c.
These plots present the results for $\mathcal{H}$ in a synthetic network and an online email network (i.e., Email-Enron), respectively, with additional results for more synthetic networks and empirical systems (High-school, Primary-school, NDC-classes) provided in the Supplementary Information. 
The results demonstrate that, at each time step, the total difference (\(\Delta \Phi^*\)) between the dynamics on hypergraph and its projected simple graph is equivalent to the sum of the dynamical uncertainty \(\Delta \Phi_1\) and the structural uncertainty \(\Delta \Phi_2\). 
Fig.~\ref{fig7:experiments}d-f presents consistent results observed in the Voter model, further demonstrating the versatility of our framework in capturing discontinuous opinion dynamics.

To systematically evaluate the generalizability of the proposed framework, we compare the differences in system instability with the sum of structural and dynamical impacts across all time steps. In a higher-order system \( \Upsilon \), we compute and normalize the area under the curve for \( \Delta \Phi^* \), \( \Delta \Phi_1 \), and \( \Delta \Phi_2 \), resulting in the metrics \( \Delta \ddot{\Phi}^* \), \( \Delta \ddot{\Phi}_1 \), and \( \Delta \ddot{\Phi}_2 \), respectively.  
The results of both DW and Voter opinion dynamics, evaluated in $\Upsilon$ across synthetic and empirical systems, are illustrated in Fig.~\ref{fig8:experiments}. The scattered points, aligned along the diagonal, highlight the generalizability and robustness of our unified framework and delineate the transitions in dynamical behavior.
These observations indicate that, akin to contagion dynamics, the transition of opinion dynamics between higher-order and pairwise networks can similarly be articulated through the additive property of structural and dynamical uncertainty.

\section{Discussion}

To conclude, we systematically study the transformability among diverse dynamics, and then identify the contributing factors for the transformation of dynamics across network orders. We reveal that structural and dynamical factors are two key drivers that account for the transformation of dynamical behaviors between dynamics on higher-order and pairwise networks. In addition, we provide a universal framework and obtain a consistent conclusion in opinion dynamics. Although we provide instances using higher-order networks with two- and three-body interactions, our framework is naturally generalized to arbitrary higher-order networks with random orders.

The fundamental transformations we establish among dynamics on hypergraphs, simplicial complexes, and dyadic graphs suggest that hypergraphs may embody structural properties of both graphs and simplicial complexes. Moreover, these relationships indicate that dynamical processes across different network representations are not independent but interconnected, revealing potential for transformation under specific conditions. 
This implies that, when studying the mechanisms of contagion dynamics, it is essential to consider not only the differences between dynamical processes but also their interrelationships.
Such insights offer valuable guidance for reconstructing real-world contagion mechanisms from observed propagation data, as previous techniques often encode the dynamics in an isolated and specific form. Additionally, the identification of this general mapping relationships paves the way for exploring the universal properties and behaviors of contagion dynamics across diverse network structures.

Moreover, this work decouples the differences between dynamics on higher-order networks and dyadic simple graphs in terms of both structural and dynamical effects. The finding enhances our understanding of how the higher-order features contribute to unique dynamical properties and behaviors during dynamical processes. These features can be conceptualized as the combined impact of multi-body interactions (i.e., structural effects) and the influence of the nonlinear nature of group interactions (i.e., dynamical effects). 
Furthermore, by scrutinizing the magnitudes of these two factors, we can discern the precise conditions where higher-order networks resist simplification. Conversely, our finding also helps determine when minimal instability loss facilitates the creation of simplified representations that preserve the integrity of the underlying dynamics.
We further generalize our findings in a unified form, and validate them across various dynamical models, including SI, SIS and SIR, as well as the DW model and Voter model in opinion dynamics, revealing the applicability and generalizability of our findings to different types of dynamics.
This acknowledges that, although dynamics on higher-order networks convey richer information than those on pairwise networks, their transitions  can  be realized  in a unified framework by exploiting the additivity of both structural and dynamical uncertainty. These findings give a crucial direction for enhancing our understanding of the dynamics of networks with higher-order structures.

This study focuses on the well-explored models in both contagion and opinion dynamics. However, a range of different models warrant further investigation, such as reactive and contact contagion processes, as well as threshold and fusion opinion dynamics, among others.
While the higher-order networks considered here are static, real-world scenarios often involve temporal group or multi-body interactions, which remains a subject for future exploration.
Moreover, our insights lay the groundwork for future research, providing a robust foundation for investigating consistent patterns across various dynamical processes, including evolutionary games~\cite{civilini2021evolutionary, xu2023evolution}, network control~\cite{de2022pinning, chen2021controllability}, and broader domains within network science~\cite{adhikari2023synchronization, von2023hypernetworks, li2024synchronization}.

\section{Methods}
\subsection{Higher-order networks}
Pairwise networks, typically modeled by simple graphs, capture interactions involving exactly two nodes, with repeated interactions between the same pair of nodes prohibited.
Higher-order networks extend the structure of pairwise networks by incorporating edges that encompass multiple nodes, referred to as higher-order interactions. These interactions capture complex, multifaceted relationships that transcend the binary links in pairwise networks.
Higher-order interactions, according to different usage scenarios, are mainly classified into two primary categories: hyperedges and simplexes. Subsequently, higher-order networks formed by hyperedges correspond to hypergraphs, whereas those constructed from simplexes are referred to as simplicial complexes. As illustrated in Fig.\ref{fig1:framework}, a hypergraph ($\mathcal{H}(V, E)$) comprises a set of nodes ($V$) and hyperedges ($E$), where each hyperedge connects an arbitrary number of nodes. Alternatively, a simplicial complex ($\mathcal{C}(V, U)$) consists of nodes, links (1-simplices), triangles (2-simplices), and a range of higher-order simplexes, where $U$ represents the collection of all simplexes of varying orders. 
Each $\delta$-simplex includes $(\delta + 1)$ nodes, emphasizing the structural hierarchy inherent in simplicial complexes.
This hierarchical organization makes simplicial complexes particularly suited for analyzing systems requiring hierarchical relationships, such as topological data analysis and high-dimensional multi-body structures, whereas hypergraphs excel in modeling group-based interactions typical of cooperative and ecological networks.

\subsection{Network projection}
Hypergraphs and simplicial complexes, both representative classes of higher-order networks, can be projected into corresponding dyadic graphs, either as multi-edge graphs (with multiple edges) or as simple graphs (without multiple edges).
The projection of a hypergraph \(\mathcal{H}(V, E)\) to a multi-edge graph involves three key steps: (i) Retain all original nodes of the hypernetwork \(\mathcal{H}\) as the node set \(V\) of the new network; (ii) For each pair of nodes, the number of edges connecting them in the new network corresponds to the number of hyperedges they share in \(\mathcal{H}\); (iii) Add the appropriate edges between node pairs to the new network. This process results in the construction of a projected multi-edge graph \( \tilde{\mathcal{H}}(V, \tilde{E})\), where \(\tilde{E}\) denotes the set of all edges in the projected network. The projection procedure for simplicial complexes is similar to this approach, by replacing hyperedges with simplices.

To convert hypergraphs and simplicial complexes into simple graphs, they are first projected into their corresponding multi-edge graphs, denoted as \(\tilde{\mathcal{H}}(V, \tilde{E})\) for hypergraphs or \(\tilde{\mathcal{C}}(V, \tilde{U})\) for simplicial complexes. In this representation, multiple edges between pairs of nodes are possible. These redundant edges in \(\tilde{E}\) (or \(\tilde{U}\)) are then removed, retaining only a single edge between any pair of nodes. The resulting set of edges, \(\hat{E}\) (or \(\hat{U}\)), defines the compressed simple graph, expressed as \(\hat{\mathcal{H}}(V, \hat{E})\) for hypergraphs and \(\hat{\mathcal{C}}(V, \hat{U})\) for simplicial complexes.

Moreover, another key projection maps a hypergraph to its corresponding simplicial complex with a mapping ratio \(\kappa\). This process compresses hyperedges larger than \(\kappa + 1\) into subsets of size at most \(\kappa + 1\), yielding a \(\kappa\)-simplicial complex. As a result, the maximum order \(\delta\) of simplices in the projected structure is constrained to \(\kappa\), ensuring that \(\max\{\delta\} = \kappa\).

\subsection{Models of dynamics}

\subsubsection{Contagion dynamics}
Contagion dynamics models are employed to illustrate the spread of viruses, information, and rumors, with the SI model being the most foundational.
In the SI model on simple graphs, states of nodes are categorized as either susceptible (S) or infected (I). At each time step, an S-state node can independently transition to the I-state upon interacting with its neighboring I-state nodes, a process quantified by the contagion rate \(\beta\). Through successive time iterations, the infected population  progressively increases until all nodes ultimately become infected.

In contrast to simple graphs, the SI model on multi-edge graphs incorporates the contagion through multiple contacts between nodes, i.e., the infection rate received by an S-state node becoming infected is heightened with the number of connections to I-state nodes, represented as \(w \cdot \beta\), where \(w\) is the number of edges linking the two nodes.

Models on higher-order networks differ from the aforementioned contagion models by highlighting multi-body interactions. 
In simplicial complexes, this is exemplified by a \(\delta\)-simplex housing a susceptible node. 
If all other nodes within this simplex are in the infected state, the susceptible node is influenced not only by direct interactions via pairwise links but also through collective propagation effects stemming from all I-state nodes. 
Specifically, in a \(\delta\)-simplex, the sole S-state node receives an infection rate denoted as \(\beta_\delta\). For \(\max\{{\delta\}} = 2\), this rate simplifies to \(2\beta_1 + \beta_2\), which are subsequently referred to as \(\beta\) and \(\beta_\Delta\) for conciseness.

Moreover, in the SI model on hypergraphs, S-state nodes transition to I-state nodes through interactions with infected nodes within their respective domain hyperedges.
The infection probability across each neighboring hyperedge is contingent upon the quantity of I-state nodes present, which can be articulated through the propagation rate \( \lambda \cdot \chi_{h}^\nu \), where \(\lambda\) and \(\nu\) are adjustable parameters, and \(\chi_{h}\) denotes the number of I-state nodes contained within each hyperedge $h (h \in \mathcal{E}_i)$ adjacent to the S-state node \(i\).

These contagion dynamics models, encompassing several network types, can also be generalized to the SIS and the SIR model. Specifically, nodes that were previously in the I-state have a defined recovery probability \(\mu\); in the SIS model, I-state nodes revert to the S-state, while in the SIR model, they transition to the R-state.

\subsubsection{Characterize contagion dynamics}

Building upon the mechanisms of contagion dynamics, we define the infection probability \(P_i(t)\) that any S-state node \(i\) in a network becomes infected at time \(t\) during the contagion process. Consequently, the contagion dynamics on a dyadic graph $\mathcal{G}$ can be articulated as \(P_{\mathcal{G}}^i(t)\):
\begin{equation}
    P_{\mathcal{G}}^i(t) = 1 - \prod_{j \in I(t)} (1 - \beta A_{ij}).
\end{equation}
Similarly, \(P^i(t)\) can be represented in the context of simplicial complex $\mathcal{C}$ and hypergraph $\mathcal{H}$ as follows:
\begin{equation}
    P_{\mathcal{C}}^i(t) = 1 - \prod_{\delta=2}^{\max{|u|}, u\in U} \prod_{j^\delta_1,\cdots, j^\delta_{\delta-1} \in I(t)} (1 - \beta_\delta A^\delta_{i,j^\delta_1,\cdots, j^\delta_{\delta-1}}),
\end{equation}
\begin{equation}
    P_{\mathcal{H}}^i(t) = 1 - \prod_{m=2}^{\max{|h|}, h\in E}\prod_{h \in \mathcal{E}_i} (1 - \lambda \chi_h^\nu A_h^m),
\end{equation}
where \(A\) denotes the adjacency matrix of the nodes, with its maximum dimension determined by the dimensionality of interactions among the nodes. For instance, in a simple graph characterized by binary edges, the maximum dimension of \(A\) is 2. In a simplicial complex where the highest order of simplices is \(\delta\), the maximum dimension of \(A\) is \((\delta + 1)\). In a hypergraph, the maximum dimension of \(A\) corresponds to the maximum number of nodes contained within any hyperedge.

To characterize the dynamical process, we employ mean-field theory. Taking the SI contagion dynamics in simple graphs as an example, if the number of edges connected to each node equals the average node degree \(\langle k^* \rangle\), the evolution of infectious nodes during the SI contagion dynamics can be represented by the following equation:
\begin{equation}
    \frac{dI_{\mathcal{G}}(t)}{dt} = (1-I(t))\cdot \left\langle k^* \right\rangle \cdot \beta \cdot I(t).
\end{equation}
For simplicial complexes, this involves extending the concept of binary edges to the simplices that are adjacent to a node:
\begin{equation}
    \frac{dI_{\mathcal{C}}(t)}{dt} = \sum_{\delta=1}^{\max{|u|}, u\in U}(1-I(t))\cdot \left\langle k_\delta^* \right\rangle \cdot \beta_\delta 
 \cdot I^{\delta}(t).
\end{equation}
In the SIS and SIR models, an additional term, namely, $-\mu \cdot I(t)$, is included in the expression, which accounts for the transition of some infectious nodes to S-states or R-states at each time step, respectively.

\subsubsection{Opinion dynamics}
Opinion dynamics models provide insights into the mechanisms underlying opinion change and consensus formation by simulating the interactions and information exchanges among individuals. This type of dynamic models helps explain phenomena like viral trends, polarization, or consensus-building in real-world social systems. Among these models, the Deffuant-Weisbuch (DW) model~\cite{dong2018survey, das2014modeling} looks at the continuous opinion changes over time. Specifically, the model's evolution is governed by interactions between nodes, where nodes only change opinions with neighbors whose opinion diversity fall within a designated threshold, termed the "confidence bound". During interactions, the opinions of individuals converge. 

On graphs, the DW model is implemented through four principal steps: (i) Randomly select a pair of nodes \( (i, j) \) from the graph. (ii) Assess whether the opinion difference of the selected pair lies within the tolerance \( \epsilon \). If the condition \((x_{i} - x_{j})^2 \leq \epsilon\) is met, the opinions of these two nodes are updated; otherwise, no change occurs. (iii) If the selected pair satisfies the compatibility condition, their opinions are updated according to the model's prescribed update rule. (iv) This process is repeated until either all nodes converge to a uniform opinion or the designated time steps conclude. In this work, the update rule involves taking the mean of the initial opinions of the interacting individuals. In scenarios involving graphs with multiple edges, the influence of these edges is incorporated such that a greater number of repeated edges \( \eta \) between two nodes leads to a reduced opinion gap, expressed as \( \frac{1}{\eta} \cdot (x_{i} - x_{j})^2 \leq \epsilon \).

Opinion dynamics can also influenced by factors like group pressure, nonlinear interactions, or higher-order effects (e.g., influence from group conversations, not just one-on-one). We then extend the DW model to hypergraphs to develop a DW opinion evolution model that accounts for higher-order interactions. In this adaptation, a pair of nodes \( (i, j) \) is selected as the initial step, but the opinion gap between them is influenced by the quantity and size of the hyperedges they share. Specifically, an increase in the number of neighboring hyperedges or the size of these hyperedges enhances the likelihood of opinion convergence. The condition for opinion updates is given by:
\[
\frac{1}{\sum_{s}|h_s|^{\alpha}}(x_{i} - x_{j})^2 \leq \epsilon, \quad (i,j) \in h_s,
\]
where \( |h_s| \) denotes the size of hyperedge \( h_s \), i.e., the number of nodes it encompasses, and \( \alpha \) is an adjustable parameter that captures the nonlinear effects of neighboring hyperedges on opinion updates.

In addition, we also introduce the multivariate Voter model with confidence bounds to study the dynamics of opinion changes~\cite{mori2019voter, shi2013multiopinion}. Unlike the DW model, the opinion values of individuals in the Voter model are multivariate and discontinuous, with opinion values ranging from 1 to 10. Another key difference from the DW model is the opinion updating mechanism: when the updating condition is satisfied, the opinion value of a randomly chosen node is updated to match the opinion value of one of its neighbors in the next time step.

\subsection{System instability}
The dynamical process represents a temporal evolution in which the state of a node changes over time steps. Taking the SI contagion dynamics as an example, the state of a node is characterized by two modes. Consequently, the probability of a node being in either the I-state or S-state can be described by the proportion of nodes in each corresponding state within the network. Thus, the information conveyed by the dynamical process can be quantified using the concept of information entropy. By measuring the chaos and instability within the system, information entropy provides a quantitative understanding of the dynamics at play. Below, the SI model is used as an example to give the information entropy measurement of the propagation process.

In the SI model, let \( |V| = N \) represent the total number of nodes in the network. At time \( t \), let \( S(t) \) and \( I(t) \) denote the number of nodes in the S-states and I-states, respectively. The probabilities of being in the S-states and I-states at time \( t \) can be expressed as follows:
\begin{equation}
    \mathbf{p}_{\text{S}}(t) = S(t) / N, \quad
    \mathbf{p}_{\text{I}}(t) = I(t) / N,
\end{equation}
Consequently, the instability in the states of susceptible and infected nodes during the dynamics of this system can be quantified by the information entropy $ \Phi(t)$ , expressed as follows:
\begin{eqnarray}
    \label{info_entropy}
    \Phi(t) =& - \mathbf{p}_{\text{S}}(t) \log_2\mathbf{p}_{\text{S}}(t) - \mathbf{p}_{\text{I}}(t) \log_2\mathbf{p}_{\text{I}}(t).
    \\ =& - (1 - \mathbf{p}_{\text{I}}(t)) \log_2(1-\mathbf{p}_{\text{I}}(t)) - \mathbf{p}_{\text{I}}(t) \log_2\mathbf{p}_{\text{I}}(t).
\end{eqnarray}
For contagion dynamics, we quantify the system's instability using information entropy, as defined in Eq.~\ref{info_entropy}. In the context of opinion dynamics, we measure \(\Phi(t)\) by assessing the dispersion of opinions across all nodes, captured by their variance, denoted as \(\mathcal{D}\).

% \section*{Acknowledgments}
% This was was supported in part by......

%Bibliography
\bibliographystyle{unsrt}  
\bibliography{references}

\end{document}